\documentclass[tightenlines, aps]{revtex4}
\usepackage{graphicx}
\usepackage{float}
\begin{document}

\title{Infrared behaviour of the pressure in $g\phi^3$ theory in 6 dimensions}

\author{ M.~E.~Carrington}
\email{meg@theory.uwinnipeg.ca}
\altaffiliation{Winnipeg Institute for
Theoretical Physics, Winnipeg, Manitoba, R3B 2E9 Canada}
\affiliation{Department of Physics, 
Brandon University, Brandon, Manitoba,
R7A 6A9 Canada}
\author{ T.~J.~Hammond}
\affiliation{Department of Physics, University of Winnipeg, Winnipeg, 
Manitoba, R3B 2E9 Canada}
\author{R. Kobes}
\email{randy@theory.uwinnipeg.ca}
\altaffiliation{ Winnipeg Institute for
Theoretical Physics, Winnipeg, Manitoba, R3B 2E9 Canada}
\affiliation{Department of Physics, University of Winnipeg, Winnipeg, 
Manitoba, R3B 2E9 Canada}

\begin{abstract}
In an earlier paper Almeida and Frenkel considered the
calculation of the pressure in $g\phi^3$ theory
in 6 dimensions via the Schwinger--Dyson equation \cite{frenkel}.
They found, under certain approximations, that a finite result
ensues in the infrared limit. We find this conclusion to 
remain true with certain variations of these approximations,
suggesting the finiteness of the result to be fairly robust.
\end{abstract}

\maketitle

\section{Introduction}
The infrared behaviour of gauge theories
at finite temperature has long been a subject of interest.
One aspect that has received particular attention is
the fact noted by Linde that the thermodynamic
pressure in Yang--Mills theory cannot be calculated perturbatively beyond
5${}^{\rm th}$ order in the coupling constant \cite{linde}. 
This problem has led
to the development of a number of ways to extend the loop expansion in gauge
theories at finite temperature, among which are approaches
based on the Schwinger--Dyson equation \cite{kk1, kk2}, 
the Braaten--Pisarski resummation scheme \cite{bp, taylor}, and
a hybrid approach of perturbation and lattice gauge theory \cite{bn}.
\par
In an earlier paper Almeida and Frenkel used the Schwinger--Dyson 
equation to study the pressure
in $g\phi^3$ theory in 6 dimensions \cite{frenkel}. This model 
is chosen because it has some qualitative similarities to Yang--Mills theory.
They showed that when a certain set of 
ladder graphs are included the pressure is finite in the
infrared limit. Given the fundamental importance of this
conclusion, it is of some interest to examine to what degree
the finite nature of the calculation is dependent on the
approximations used. The purpose of this note is to
study this question.
\par
The paper is organized as follows. In Section \ref{af}
we set up a general framework for the calculation of the
self--energy in the model, and show what approximations
are used within this framework to obtain the results of
Almeida and Frenkel \cite{frenkel}. Section \ref{other}
contains a discussion of four related variations of these
approximations, which differ basically by including different 
approximations to the full propagator
and vertex functions. We find that these different approximations
lead to a quantitative difference with the results of 
Almeida and Frenkel \cite{frenkel}, but the conclusion of the
finite nature of the result in the infrared limit remains the
same. Section \ref{end} contains some brief concluding remarks.
\section{Almeida--Frenkel Approximations}
\label{af}
The pressure in a field theory can be calculated by
finding the self--energy function and using the fundamental
relation:
\begin{equation}
  \left( \frac{\delta P}{\delta {\cal D}_0} \right) =
-\frac{T}{2} \Pi
\end{equation}
where ${\cal D}_0$ is the free propagator and only
one--particle--irreducible graphs are included in the
self--energy function $\Pi$. To calculate this self--energy 
for $\phi^3$ theory in 6 dimensions 
we consider 
the Schwinger--Dyson equation:
\begin{equation}
\Pi(p) = \frac{1}{2} \frac{g^2T}{(2\pi)^5}\int_\lambda^T
d^5k\ \Gamma(p, k) G(k) G(p+k),
\label{sd}
\end{equation}
where $\lambda$ is an infrared cutoff,
$G^{-1}(p)=p^2+\Pi(p)$ is the full inverse propagator, 
$\Gamma(p,k)$ is the full 3--point function, and
we have taken the high temperature limit so that
only the $n=0$ term in the Matsubara frequency sum contributes.
\par
Consider now the approximation
\begin{eqnarray}
 && G(p) \approx
\frac{1}{p^2} -\frac{\Pi(p)}{p^4}\nonumber \\
&&\Gamma(p,k) \approx 1
\label{fra}
\end{eqnarray}
Under this assumption, Eq.~(\ref{sd}) becomes
\begin{equation}
  \Pi(p) = \frac{g^2T}{48\pi^5p^3} \left\{\frac{1}{2}
\int_\lambda^T k\,dk\ \log\left(\frac{k+p}{k-p}\right)^2
-\int_\lambda^T k\,dk\ \log\left(\frac{k+p}{k-p}\right)^2
\frac{\Pi(k)}{k^2}\right\}
\label{frenkel}
\end{equation}
If we now make the further approximations
\begin{eqnarray}
  \label{log}
  \log\left(\frac{k+p}{k-p}\right)^2 
  &\approx& \frac{4k}{p}\qquad\qquad \ldots k \ll p \nonumber\\
  &\approx& \frac{4p}{k}\qquad\qquad \ldots k \gg p
\end{eqnarray}
and define
\begin{eqnarray}
  \label{defns}
  \alpha &=& \frac{g^2}{12\pi^3}\nonumber\\
  x &=& \frac{\alpha T}{p}\nonumber\\
  y &=& \frac{\alpha T}{k}\nonumber\\
  f(x) &=& \frac{\Pi(p)-p^2}{p^2}
\end{eqnarray}
Eq.~(\ref{frenkel}) becomes
\begin{equation}
  \label{eq:frenkel}
  f(x) = -1+\frac{x}{3}-\frac{x^2}{2\alpha}-
  x^2\int_\alpha^{x}\frac{dy}{y^2}f(y)
  -x^4\int_x^{\infty}\frac{dy}{y^4}f(y).
\end{equation}
This equation was derived by Almeida and
Frenkel \cite{frenkel}, who show, analytically, that the
solution for $f(x)$ is finite in the infrared limit
$x\to \infty$. This result demonstrates that, using the 
approximations discussed above, the pressure is finite in the infrared limit. 
\section{Other approximations}
\label{other}
In this section we explore approximations other than
Eqs.~(\ref{fra}),~(\ref{log}) in solving the Schwinger Dyson equation
(\ref{sd}). We consider two distinct questions: what happens
if we don't make the approximation of Eq.~(\ref{log})
for the logarithmic terms, and what happens if we use a
non--trivial 3--point interaction. As these approximations
will require a numerical solution, in the following we choose
$\alpha=3$.
\subsubsection{Approximation A}
\begin{figure}[H]
\begin{center}
\includegraphics[height=7cm]{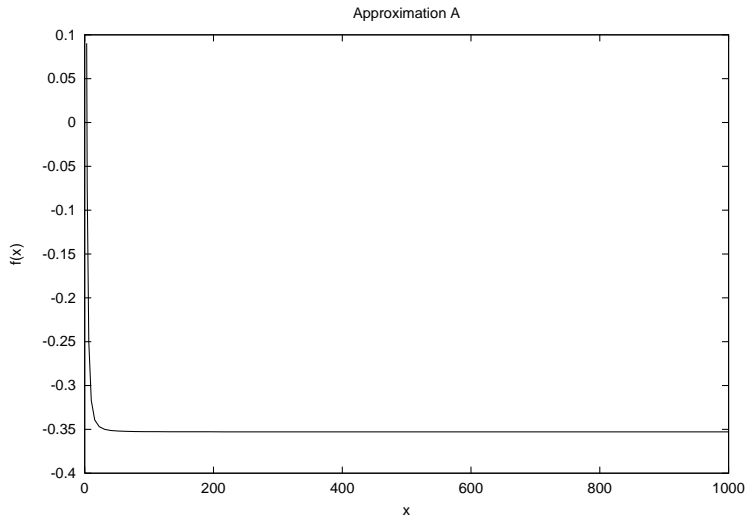}
\end{center}
\caption{Approximation A}
\label{afig}
\end{figure}

We first explore what happens
if we don't make the approximations of Eq.~(\ref{log}) to the
logarithmic term of Eq.~(\ref{frenkel}).
In this case, Eq.~(\ref{frenkel}) can be written in the form
\begin{equation}
  \label{eq:A}
  f(x) = -1-\frac{x}{16\alpha^2}\left[
    (x^2-\alpha^2)\log\left(\frac{x+\alpha}{x-\alpha}\right)^2
    +4\alpha x\right]
  - \frac{x^3}{4}\int_\alpha^{\infty}
  \frac{dy}{y^3} \log\left(\frac{x+y}{x-y}\right)^2 f(y)
\end{equation}
in the infrared limit $\lambda \to 0$.
Solving this equation numerically gives the same results
as that found in \cite{frenkel} for the analytic solution
of Eq.~(\ref{eq:frenkel}). These
results appear in Fig.~\ref{afig}.
\par
\par
The integral equation (\ref{eq:A}), and the equations that will be 
considered below, are subtle to solve, especially
in the limit of large $x$ (small momentum), which is the region we
are mainly interested in. The source of the difficulty is as follows. 
In the limit $x\to\infty$ Eq.~(\ref{eq:A})  
has the form
\begin{equation}
  f(x) = g(x) + x^3\int_\alpha^\infty\frac{dy}{y^3}K(x,y) f(y),
\end{equation}
This equation has a contribution from $g(x)\sim x^2$
which is canceled by a term coming from the integral. This cancellation is 
difficult to see numerically, but it can be avoided by performing 
the substitution,
\begin{equation}
  f(x) = \tilde f(x)+\frac{\alpha^2\sigma}{x^2} +f_\infty
\end{equation}
with constants $\sigma$ and $f_\infty$ to be fixed shortly.
Using this substitution, we find that $\tilde f(x)$ satisfies an equation
of the form
\begin{equation}
  \tilde f(x) = \tilde g(x,\sigma,f_\infty) 
  -\frac{\alpha^2\sigma}{x^2} -f_\infty
  + x^3\int_\alpha^\infty \frac{dy}{y^3}K(x,y) \tilde f(y)
\end{equation}
We fix $\sigma$ by requiring $\tilde g$ approach a
constant as $x\to\infty$, and we fix $f_\infty$ by requiring
the solution $\tilde f(x)$ approach 0 as $x\to\infty$.
\subsubsection{Approximation B}
\begin{figure}[H]
\begin{center}
\includegraphics[height=7cm]{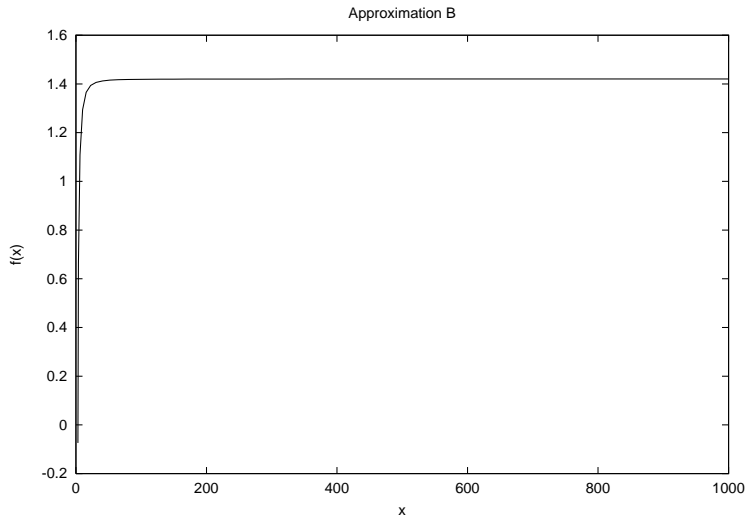}
\end{center}
\caption{Approximation B}
\label{bfig}
\end{figure}

In this section and the next we use an eikonal approximation for the
3--point function \cite{cornwall, hou, us}:
\begin{equation}
  \label{eikenol}
  \Gamma(p,k) \approx 1 + \frac{\Pi(k+p) - \Pi(k)}
  { {\vec p}\,^2 + 2 {\vec k}\cdot {\vec p} }
\end{equation}
Using this expression effectively resums a series of higher 
loop vertex graphs which are important in the infrared region.
First, we consider the integral equation obtained from Eq.~(\ref{sd}) 
with the vertex given by Eq.(\ref{eikenol}) and bare propagators:
\begin{eqnarray}
  G(p) &\approx& \frac{1}{p^2} 
\label{appb}
\end{eqnarray}
Using the notation
of Eq.~(\ref{defns}) and taking the limit $\lambda\to 0$ we obtain,
\begin{eqnarray}
  \label{eq:B}
  f(x) &=& -1+\frac{x}{4\alpha^2}\left[
    \frac{3}{4}(x^2-\alpha^2)\log\left(\frac{x+\alpha}{x-\alpha}\right)^2
    -\frac{1}{8}(4x^2-\alpha^2)\log\left|\frac{2x+\alpha}{2x-\alpha}\right|
    +\frac{5}{2}\alpha x\right] \nonumber\\
  &-& \frac{x^3}{4}\int_\alpha^{\infty}
  \frac{dy}{y^3}\left\{ \frac{3}{2}\log\left(\frac{x+y}{x-y}\right)^2 
    - \log\left|\frac{2x+y}{2x-y}\right|\right\}f(y)
\end{eqnarray}
 The results, which appear in Fig.~\ref{bfig}, indicate that
the self--energy function is finite in the infrared
limit $x\to\infty$, although, not surprisingly, we obtain an
asymptotic value quantitatively different from that
of Fig.~\ref{afig}.
\par
\subsubsection{Approximation C}
\begin{figure}[H]
\begin{center}
\includegraphics[height=7cm]{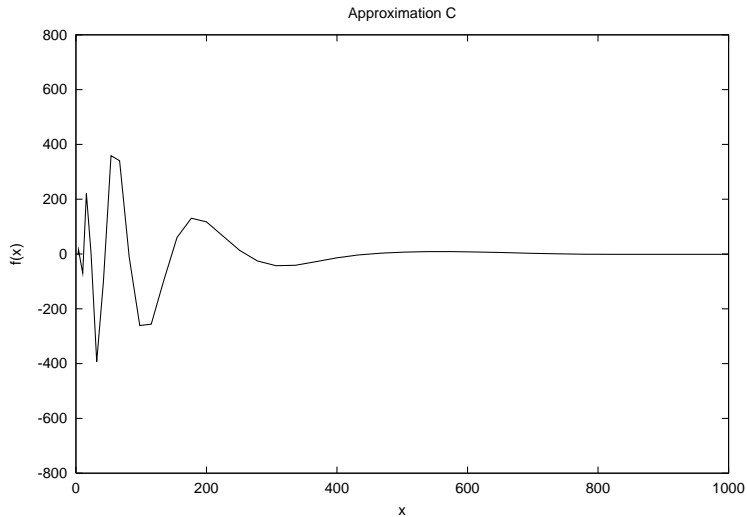}
\end{center}
\caption{Approximation C}
\label{cfig}
\end{figure}

We next consider the integral equation obtained from Eq.~(\ref{sd}) using
\begin{eqnarray}
  G(p) &\approx& \frac{1}{p^2} -\frac{\Pi(p)}{p^4}\nonumber \\
\Gamma(p,k) &\approx&  1 + \frac{\Pi(k+p) - \Pi(k)}
  { {\vec p}\,^2 + 2 {\vec k}\cdot {\vec p} }
\label{appc}
\end{eqnarray}
Using the notation
of Eq.~(\ref{defns}) and taking the limit $\lambda\to 0$ we obtain,
\begin{eqnarray}
  \label{eq:C}
  f(x) &=& -1+\frac{x}{4\alpha^2}\left[
    \frac{1}{4}(x^2-\alpha^2)\log\left(\frac{x+\alpha}{x-\alpha}\right)^2
    -\frac{1}{8}(4x^2-\alpha^2)\log\left|\frac{2x+\alpha}{2x-\alpha}\right|
    +\frac{1}{2}\alpha x\right] \nonumber\\
  &-& \frac{x^3}{4}\int_\alpha^{\infty}
  \frac{dy}{y^3}\log\left|\frac{2x+y}{2x-y}\right|f(y)
\end{eqnarray}
 These results appear in Fig.~\ref{cfig}. As before, a finite result in
the infrared limit is obtained. The fluctuations that occur at 
small $x$ are not physical since small $x$ corresponds to the ultraviolet
limit, and the eikonal approximation
of Eq.~(\ref{eikenol}) is only valid in the infrared limit.
\par
\subsubsection{Approximation D}
\begin{figure}[H]
\begin{center}
\includegraphics[height=7cm]{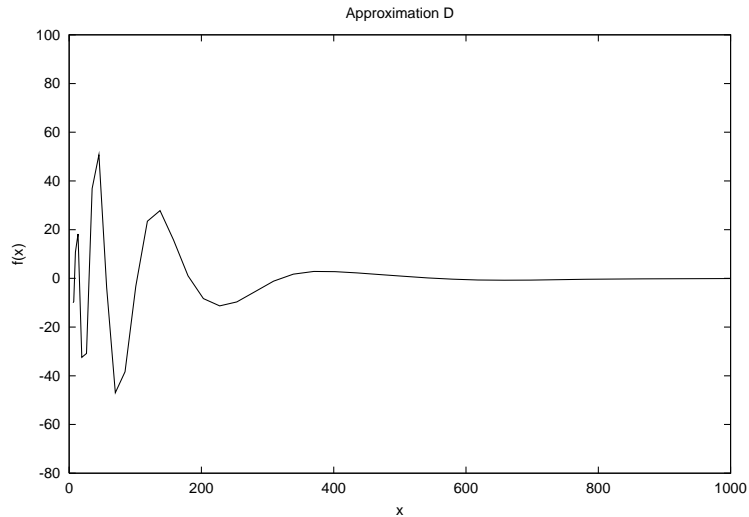}
\end{center}
\caption{Approximation D}
\label{dfig}
\end{figure}

The final approximation we consider is a variation of the previous
one. We use the eikonal expression Eq.~(\ref{eikenol}) for the vertex 
and work with full propagators of the form: $G(p) = 1/p^2+\Pi(p)$. 
The Schwinger--Dyson equation becomes: 
\begin{equation}
\Pi(p) = \frac{g^2T}{(2\pi)^5}\int_\lambda^T
d^5k\ \frac{ G(k)}{ {\vec p}\,^2 + 2 {\vec k}\cdot {\vec p} }
\end{equation}
At this point we make the approximation
\begin{equation}
   G(p) \approx \frac{1}{p^2} -\frac{\Pi(p)}{p^4}
\end{equation}
and obtain, in the limit $\lambda\to 0$,
\begin{equation}
  \label{eq:D}
  f(x) = -1-\frac{x^3}{4}\int_\alpha^{\infty}
  \frac{dy}{y^3}\log\left|\frac{2x+y}{2x-y}\right|f(y)
\end{equation}
The difference between Eq.~(\ref{eq:D}) and Eq.~(\ref{eq:C}), which 
was obtained using Approximation C, is as follows. Eq.~(\ref{eq:D}) 
makes use of an analytic cancellation between certain sets of graphs 
corresponding to vertex corrections, which are contained in 
Eq.~(\ref{eikenol}), and graphs that are contained in the expression 
for the full propagator $G(p)$ \cite{us}. These analytic cancellations 
are performed, before the expansion of the full propagator is done.
The results appear in Fig.~\ref{dfig}.  Once again we obtain 
a self--energy that is finite in the infrared limit.

\par
\section{Conclusions}
\label{end}
We have considered various approximations to the
Schwinger--Dyson equation that can be used to calculate the pressure
in $g\phi^3$ theory in 6 dimensions. The approximations
used  sum up different classes of diagrams
believed to be important in the infrared limit. In all cases
considered a finite result is obtained for the pressure in
the infrared limit, although the quantitative behaviour
differs according to which approximation is used, as is
expected. The important conclusion to be drawn from these results 
is that the finiteness of the pressure 
in the infrared limit is not very sensitive to the particular
approximation used, and thus seems to be a general feature 
of pressure calculations based on a Schwinger--Dyson 
approach.
\acknowledgments
This work was supported by the Natural Sciences and Engineering
Research Council of Canada.

\end{document}